Pasquale Tucci
A Cassini 4[th] pamphlet against Delambre on lunar libration and other issues

Abstract
In this paper I analyze the context in which Cassini 1[st] described lunar libration and proposed its interpretation.

## 1.      Introduction

Between 1822, the year of death of Delambre (1749-1822), and 1827, the year of the publication of Delambre's *Histoire de l'Astronomie au dix-huitième siècle*, Jean-Dominique Cassini (Cassini 4[th]) (1748-1845), great-grandson of Jean-Dominique Cassini (Cassini 1[st]), (1625-1712) published a pamphlet. In it, quite short (16 pages), undated, Cassini 4[th] highlighted several inaccuracies of Delambre's *Histoire* were. The *Histoire* was edited, with some additions and revisions, by Claude Louis Mathieu, based on an almost complete manuscript left by the late author.

Cassini 4[th], after expressing its appreciation for the enormous work done by Delambre to make known rarely cited works in the various books, noted that in the *Histoire de l'Astronomie au dix-huitième siècle* Delambre had not followed the same criterion used in his other books on the History of Astronomy already published. It is true that in the manuscript of the *Histoire* Delambre, according to Cassini 4[th], showed vast and profound knowledge, great tact, impartiality but

> … malheureusement il n'a pas eu le temps de lire et relire sono Ouvrage. … Le manuscrit a été remis à un éditeur très-digne de cette confiance, mais qui a cru devoir de faire imprimer dans presque toute son intégrité. (Cassini Jean-Dominique, Cassini 4[th], p. 3)

In the *Histoire*, in fact, there were unfair prejudices, risky interpretations and several inaccuracies. One of them concerns Cassini 1[th] lunar libration; I will dwell on this point. But before analysing Cassini's 4[th] criticism of Delambre I will expose Cassini's (1[th]) contribution to lunar libration and Cassini's (2[nd]) insights. Next I will outline the scientific context in which Delambre matured his criticism of Cassini 1[th].

## 2.      Cassini 1[st] on Lunar libration

It was long been known that the Moon always shows the same face to the Earth. Galileo and then Hevelius had highlighted with telescopic observations that in fact a terrestrial observer observes more than 50% of the lunar surface.

If Moon's motion of revolution around the Earth were perfectly circular and if Moon's rotation took place in the same amount of time as its motion of revolution around the Earth, an observer located in the centre of the Earth should see only half of the lunar surface. But the Earth has a motion of rotation, revolution, precession and nutation and the Moon, as well as the other planets, travels along an elliptical orbit at variable speed. In addition, the motion of lunar rotation is uniform. It follows that the terrestrial observer observes more than 50% of the lunar surface.

The phenomenon was described and interpreted by Cassini 1[st] in the Memoir *De l'origine de l'Astronomie et de son usage dans la geographie et dans la navigation*, published in 1693. In paragraph 6, Cassini 1[st] reflected on why astronomers had observed significant variations in sunspots, whereas for the Moon hypotheses were made such as that the differences we see in the position of the lunar spots depended on the different way in which they are illuminated by the Sun. Cassini 1ˢᵗ instead resorted to an explanation of the phenomenon "très simples & très naturelle":

> Comme les Coperniciens attribuent deux mouvements à la terre, l'un annuel & l'autre journalier; de même on a considéré dans la lune deux mouvements différents. Par l'un de ces mouvements dont la révolution s'achève en 27 jours 8 un tiers, la lune paroît [paraît] tourner d'orient en occident sur un axe parallèle à celui de son orbite. L'autre mouvement se fait réellement d'occident en orient sur un axe dont les pôles sont éloignés de ceux de l'orbite de la lune transportée dans son globe de sept degrés & démy [demi], & des pôles de l'écliptique, de deux degrés & démy [demi]; & il a pour colure ou premier méridien le cercle de la plus grande latitude de la lune transporté aussi dans fon globe. De la complication de ces deux mouvements contraires, dont l'un n'est qu'apparent & l'autre est réel, l'un est inégal & l'autre égal, résulte la libration apparente de la lune. (Cassini 1ᵗʰ 1693, pp. 34-35)

Librations mentioned by Cassini 1ˢᵗ were apparent, as they did not depend upon any actual inequality in the motion of the Moon around its axis.

It is worth emphasizing two points in Cassini's quotation:

a) Cassini 1ᵗʰ was not only interested in the description of the phenomenon but also in its interpretation according to the hypothesis of the Copernicans.
But his position was very cautious: he used the authority of the Copernicans to support his hypothesis but on the other hand the movements attributed to the Earth did not concern its revolution around the Sun, something that for Copernicans was out of the question;

b) Cassini 1ᵗʰ, perhaps was the first to explicitly hypothesize that the Moon rotated around its axis.

Sixteen years earlier, in 1677, Cassini 1ˢᵗ had realised that Moon's trajectory was between two ovals shown by dots in (Fig.1). Cassini then used Kepler's elliptical model and showed how Rudolphine Tables were able to provide the movements of the Sun, the Moon and its apogee and so on.

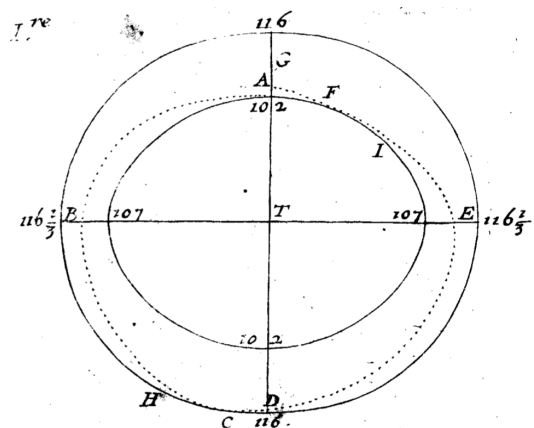

Fig. 1. Cassini's ovals, showing the Moon positions relative to the Sun in various configurations.

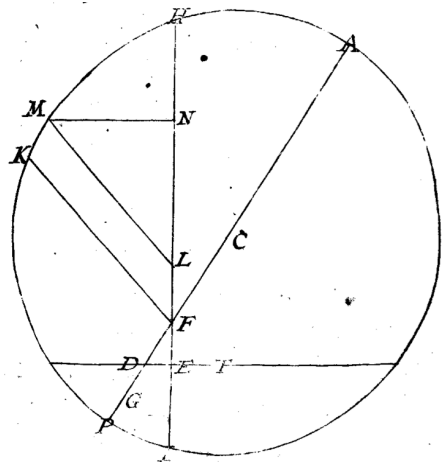

Fig. 2 Kepler's ellipse (AP as the major axis), as depicted in Cassini's 1677 memoir.

When Cassini presented his oval hypothesis, the contrast with Kepler's ellipse was not very widespread among astronomers. It was only after the derivation of Kepler's laws from Newton's theory of gravitation that Kepler's hypothesis gained the upper hand. Difference between the oval and the ellipse in the description of the trajectory of a planet was very minimal. Really Kepler's ellipse and Cassini's oval are barely distinguishable when orbits with a small eccentricity are considered. Fig. 3.

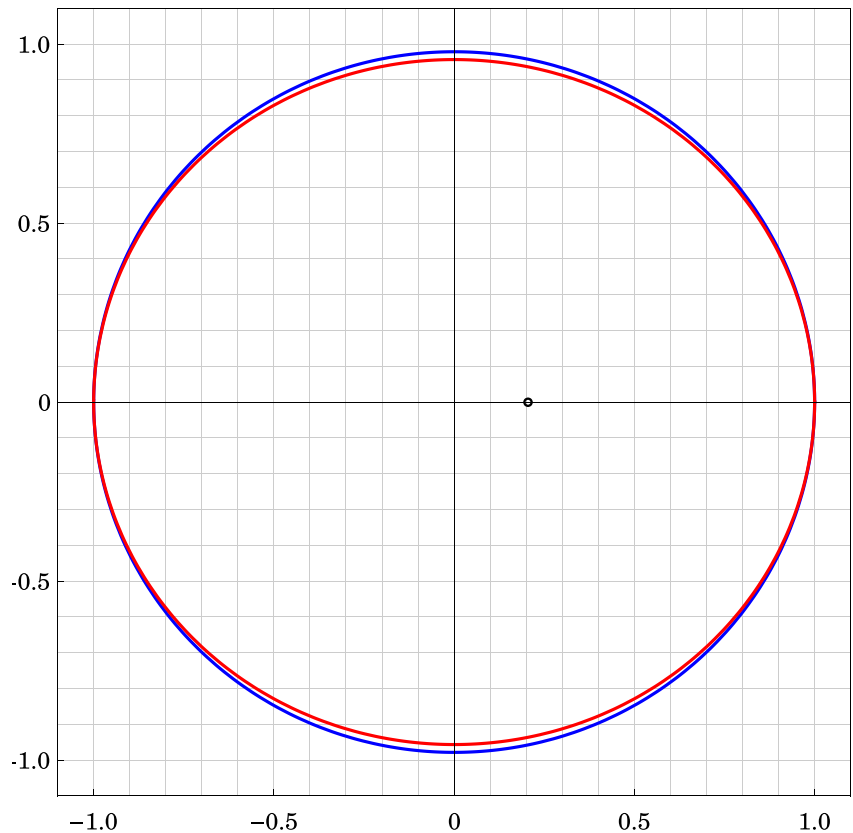

Fig.3 A blue Kepler's ellipse and a red Cassinian oval are plotted with Mercury's parameters.
(Morgado & Soares 2014, p. 6)

But the little difference concealed a decisive argument for the acceptance of elliptical instead of oval motion: according to the Newtonian theory of gravitation no central force could be responsible for the motion of a planet on a Cassini oval. (Sivardière 1994, p. 64)

In 1656 Cassini 1st had proved the validity of Kepler's second law by observing the image of the Sun at the sundial of S. Petronio. Newton had not yet published his *Principia* at that time.

Cassini 1st semi-quantitative considerations on lunar libration, translated in three "Cassini's laws" by Tisserand (Tisserand 1891), were the result of decades of observation of the lunar surface.

In 1677 Cassini 1st published "Nouvelle théorie de la Lune" and in 1679 Cassini published a map of the Moon (See Fig. 5) based on hundreds of drawings made between 1671 and 1679. The more than sixty drawings were assembled in a notebook by Cassini 4th. (Fig. 6) Moreover Cassini 1st published several articles on Moon's eclipses, which were very important for detailed lunar observations.

As reported by du Hamel, the first secretary of the Académie royale des sciences de Paris, Cassini had presented his hypothesis on lunar libration on September 7, 1675.

Scientific lives of Cassini 1st (1625-1712) and of Isaac Newton (1642-1726) overlapped for many decades. The English scientist knew many of Cassini's findings, but probably hadn't read Cassini 1th description of lunar libration , currently known, from Tisserand onwards, as "Cassini laws". On the other hand, Cassini was not interested in studying Newton's *Principia*. Newton derived Kepler's laws from his Theory of Gravitation, and this provided a new context for the study of lunar motions.

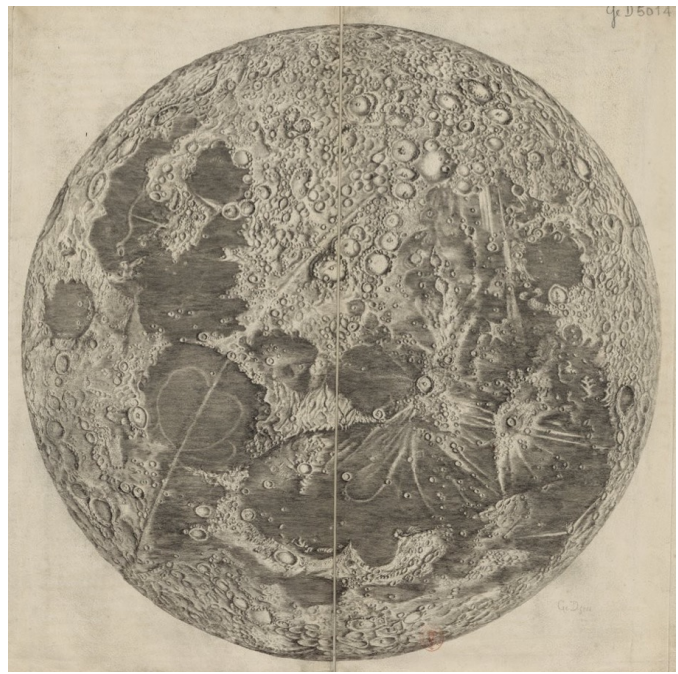

Fig. 5 Jean Dominique Cassini (Cassini 1st), Carte de la Lune, Paris, 1680. 53x53 cm
Bibliothèque nationale de France

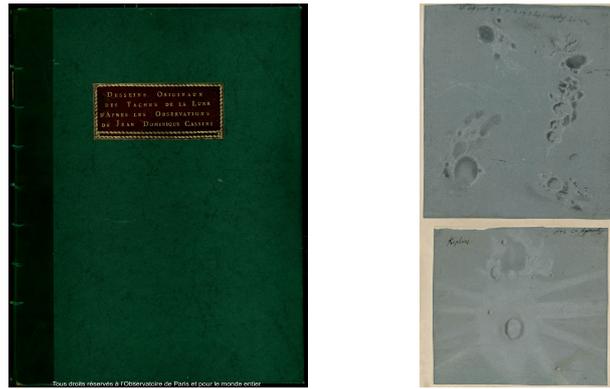

Fig. 6 "Dessins originaux des taches de la Lune"

Cassini's semi-quantitative considerations on lunar libration were developed by Jacques Cassini (Cassini 2nd), his son, in a Memoir read at the Académie royale de Sciences de Paris on June 21 and July 5, 1721, and published in 1723 (Cassini Jacques 1723). J. Cassini exposed the lunar libration in *Élémens d'Astronomie* published in 1740 (Cassini J. 1740a) and, in the same year, published the Tables of the Moon, together with those of the Sun, the Planets, the Fixed Stars and the satellites of Jupiter and Saturn. (Cassini J. 1740b)

The Memoir published in 1723 emphasized the importance of his father's contribution to the problem of libration that some astronomers, not believing in the rotation of the Moon on its axis, attributed to some compensatory motion, as though a sphere were shifting its centre of gravity.

His father attributed the libration of the Moon to the combination of a rotation motion on its axis and revolution around the Earth. Cassini 1st based his theory on the following data:

1. The equatorial plane of the Moon was inclined by 2.5 on the ecliptic and 7.5° with respect to the orbit of the Moon.
2. The rotation of the Moon around its axis took place in $27^d\ 5^h$ and was the same as that taken by the Moon to return to the initial node, i.e. to return to the point of intersection with the ecliptic.
3. The axis of rotation of the Moon, inclined by 2.5° with respect to the ecliptic, described a double cone and took 18 years and 7 months to return to its initial position. That time was equal to the time it took for a lunar node to return to its initial position.

The motion of the lunar axis was therefore completely analogous to that hypothesized by Copernicus for the Earth's axis, which takes about 25.000 years to return to its initial position. In fact, we know that the Earth's axis also has a nutation motion, i.e. oscillation around its circular trajectory, discovered by Bradley after Cassini's semi-quantitative considerations on Moon's libration.

Jacques Cassini then took into consideration the movement of the fixed stars with respect to the Moon, the libration movement of the Moon with respect to the fixed stars, the libration movement of the Moon with respect to the Sun. This last case is relevant because J. Cassini showed that it is the Moon that rotated around its axis, otherwise we would have to admit that it was the Sun that made a motion of revolution around the Moon, which was completely unimaginable.

Ultimately, libration is caused by the rotation of the Moon around its axis in combination with the movement of revolution around the Earth in the opposite direction.

3.　　Gravitational interactions after Newton and discussions about the law of the inverse square of distance

With the diffusion of the Newtonian theory of gravitation, it was possible not only to derive from it the three laws of Kepler, but it was necessary to face the problem of the gravitational interaction between the Earth, the Sun and the Moon. The two-body problem had already been faced and solved by Newton in the hypothesis of two point-masses that attract each other according to the law of the inverse square of the distance.

If we also consider the attraction exerted by the Sun, we are in a field where the Newtonian theory seemed to be wavering. But not even the attraction exerted between three bodies was sufficient to give a complete explanation of inequalities of Moon's trajectory that were recorded by astronomical observations. Further, more precise observations pointed to irregularities that seemed to call into question Newtonian theory.

Starting in the '40s of the eighteenth century, Euler (Euler 1749); (Euler 1753); Clairaut (Clairaut 1752), and d'Alembert (d'Alembert 1754) approached the three-body problem in the context of Newton's theory of gravitation. In the specific case of the Moon, the authors took into consideration not only the gravitational force exerted by the Earth but also that of the Sun, through the Leibnizian calculus. The problem had also relevant practical implications. Since the establishment of the first astronomical observatories in London and Paris lunar theory and reliable lunar position tables, seemed to offer a possible method for finding longitude at sea.

Clairaut in his *Théorie de la Lune* published in 1752 was the first who published a lunar theory which gave explicit derivation of the inequalities of Moon's motion by means a Leibnizian calculus. Previously, both Clairaut and d'Alembert and Euler had discovered, independently and through different mathematical methods, that their calculations gave a result in disagreement with the observations for the motion of the apsis of the Moon.

Clairaut in 1747 had proposed that a term that varied inversely to the fourth power of distance had to be added to the law of the inverse square of distance.

Euler, on the other hand, thought that the inverse law of the distance was not applicable to large distances and hypothesized a different role of the aether depending on the distance between the gravitationally interacting bodies.

d'Alembert, on the other hand, argued that the law of the inverse square of the distance was correct and that the cause of the discrepancy between observational data and theory in the case of the motion of the lunar apsis came from a different force, such as a magnetic force, which acted together with the gravitational force.

The question was settled when Clairaut, instead of stopping at the first order of approximation, continued the series development until the approximation of the second order. However, the predictive accuracy in the lunar theories was comparable to that which could be obtained from the Newtonian Theory that Le Monnier had published in 1746 in *Institutions astronomiques*. (Keill, 1746) The first lunar tables accurate enough to give position of the Moon to two arc-minutes, and hence to give navigators the geographical longitude to 1°, were those of Tobias Mayer (1723-1762), published in 1753. (Wilson 2010, p. 12)

Mayer's tables were semi-empirical and did not answer the theoretical question as to whether the Newtonian law could account for all lunar inequalities. But they met navigator's practical need. It was later replaced by the chronometric method. However Astronomical method was still used for a long time because it provided an important check on chronometrical method. Mayer had also corrected 2.5° to 1.25° in Cassini semi-quantitative considerations about lunar libration.

In 1778 Charles Mason revised Mayer's tables, relying on 1137 observations due to Bradley. It was in the same way, apparently, that Tobias Bürg revised Mason's tables using 3000 of the Greenwich lunar observations made by Maskelyne between 1760 and 1793. But the accuracy of the tables depended crucially on the empirical refining of constants. (Wilson 2010, p. 13)

Laplace's theory (Mécanique Céleste, Book VII, Introduction) was considerably more accurate than the earlier analytical theories of Clairaut, Euler, and d'Alembert. But theoretical deduction fell little short of the accuracy attainable by comparisons with observations.

In 1818 P. S. Laplace (1749-1827) was able to claim confidently that the problem had been solved from both a theoretical and a practical (compilation of tables) point of view for all the bodies in the solar system; among those that still required better and more accurate tables, only the Moon remained. Laplace's claim was in large part self-commendatory for the decisive contribution that he knew he had made to the problem so it may give the false impression that a solution had been reached by following a line which led straight to the conclusion contained in his *Mécanique Céleste* (1799) in which he summarized and gave organic shape to the theory he had been working on in the previous two decades. (Tagliaferri & Tucci 1999, p. 222 )

In 1818, Laplace proposed that the Académie des Sciences set up a prize to be awarded in 1820 to whomever succeeded in constructing lunar tables based solely on the law of universal gravity. The announcement of the prize was worded as follows:

> Former, par la seule théorie de la pesanteur universelle, et n'empruntant des observations que les élémens arbitraires, des tables du mouvement de la lune aussi précises que nos meilleurs tables actuelles.

The prise was awarded to Carlini-Plana and Damoiseau by a commette of which Laplace was a member.

4.      Delambre on lunar libration of Cassini 1st

After the paragraph on the context of Delambre critiques, I come back to the Jean-Dominique Cassini (Cassini 4th).

Delambre had written:

> … mais sur quels fondemens a-t-il appuyé sa théorie? C'est ce que nous avons inutilement cherché jusqu'à ce jour, et qui nous a fait dire que Dominique [Cassini 1st] n'a fait que copier Kepler. (Mathieu (ed) 1827, p. 265)

Delambre referred to the theory read at the meeting of the Académie des sciences on June 21 and July 5, 1721, and published in (Cassini (Cassini 2nd) 1723). Jacques Cassini re-proposed the Theory of Lunar libration, according to the theory exposed by Cassini 1st, in the Livre III of *Élémens d'Astronomie*. (Cassini [Cassini 2nd] 1740a, pp. 255-334.)

Cassini 4th opposed to Delambre that while he justified Lacaille, who in 1742 had used a method that had been previously expounded by Kepler, stating that it was correct to resort to the work of others, for Cassini 1st he didn't use the same criterion. And Cassini 4th commented:

> M. Delambre ne s'aperçoit pas que *l'animosité* trop marquée avec laquelle il poursuit D. Cassini doit ôter toute confiance dans les jugements qu'il porte sur lui. (Cassini J.D. (Cassini 4th) ----, p. 8)

Cassini 2nd, according to Delambre, belonged to an "école de Cassini" which
> … en general calcule très peu; elle ramène tout à des constructions graphiques … en traitant cette matière dans le seul but de ne point s'écarter des idées de Dominique [Cassini 1st] et de soutenir le système de Descartes." (Delambre 1827, p. 254)

Delambre criticized the Cartesians specifically about the shape of the Earth and the theory of vortices applied to comets.

Delambre, in *Astronomie moderne*, had included a discussion of Cassini 1st lunar libration along with an analysis of Cassini 1st entire scientific output. The judgment on lunar libration is negative:

> L'explication de Cassini [Cassini 1st] aurait pu être exposée d'une manière plus lumineuse. (Delambre 1821, Tome second, p. 734) Elle est détaillée en deux pages et demie dans les élémens di J. Cassini (Cassini [Cassini 2nd] 1740a, pp. 255-260), sans y devenir plus claire." (Delambre 1821, p. 734)

Graphical method used in Cassini's research allowed at most the appreciation of half a degree, while the new methods of analysis required much more precise data.

J. Cassini (Cassini 2nd) returned to the subject in Livre III of his *Elémens d'Astronomie* (Cassini 1740a, pp. 251-334). The book, published in the same year, also containing the Astronomical Tables of the Moon, must be linked to it. (Cassini 1740b)

The Cassini 2nd Tables were complicated, according to Delambre, using ellipses and epicycles. Such use of graphic elements could have been remedied with analytical methods. The tables, moreover, were inaccurate and any artifice to make them more precise would have required a continuous effort to adapt them. (Delambre 1827, pp. 265-266) In this regard, Delambre recalled how Cassini de Thury (Cassini 3rd), in a book of 1764, corrected the Tables of the Moon provided by his father Jacques Cassini. (Cassini 3rd 1764)

Delambre's judgment on Jacques Cassini's book was critical. According to Delambre, Jacques limited himself to exposing his father Jean-Dominique's System of Knowledge, thus remaining a hundred years behind the knowledge acquired.

Although Jacques Cassini remained a Cartesian at a time when Newtonianism was gaining ground, Delambre recognized that Jacques Cassini's astronomical observations could still be used for useful calculations. And he concluded with a praise of Cassini 2nd:

> Quand J. Cassini [Cassini 2nd] a suivi une fausse route, il n'a égaré personne. Il lui reste ses travaux utiles pour la longueur de l'année, pour la diminution de l'obliquité, et toutes ses observations de détail qui sont en grand nombre. (Delambre 1827, p. 274)

Cassini 4th resigned on 6 September 1793. We are at the beginning of the Régime de la Terreur, a difficult period in which the events of the appointment of the new director were inserted. (Chapin 1990) The Académie had designated Cassini 4th, along with Legendre and Mèchain, to measure the meridian arc. Legendre had declared that he did not accept; Cassini wrote to the Académie that he could not take charge of it. According to Delambre, he resigned as he wanted nothing to do with the new authorities of the time with whom he deeply disagreed.[i] As soon as Cassini 4th resigned, it received the order to leave the Observatory in 24 hours. The following year he was imprisoned; his detention lasted seven and a half months. Jean Perny de Villeneuve was appointed interim director; he was succeeded by Joseph Jérôme de Lalande.

Delambre acknowledged Cassini 4th, when he was director of the Astronomical Observatory from 1784 to 1793, for having committed himself to the improvement of the structure, now in very poor condition, without instruments and without personnel. (Cassini 4th 1810, p. VI; Wolf 1902, p. XI; Chapin 1990)

However, despite very polite tones, disagreement between Cassini 4th and Delambre was deep and concerned the way of doing science. The contrast had been evident ever since several mathematicians and astronomers had tried to deduce lunar motion from Newton's theory of gravitation.

## 5.    Delambre's scientific achievements

In 1780 Delambre had begun to attend the lectures of Lalande at the Collège de France; eventually he became his assistant and scientific collaborator. (Cohen 1975) In 1784 Delambre attended the session of the Académie des Sciences when Laplace read his theory of the great Inequalities of Jupiter and Saturn (Laplace 1788a) and Laplace 1788b).

Laplace perfected the Theory of the Satellites of Jupiter in a memoir published in 1787 (Laplace 1787) and in a memoir, in two parts, for 1788 and 1789, published respectively in 1791 and 1793 (Laplace 1791 and 1793). In his *Mécanique Céleste* all these problems were tackled in a comprehensive way.

While Laplace was perfecting his theory, Delambre analysed almost all ancient and modern observations and published several Tables in the book *Astronomie* of Lalande, third edition, published in 1792. One of them dealt with the eclipses of the four satellites of Jupiter. In the explanation of the Table Delambre stated

> La théorie des attractions mutuelles des satellites, donnée par M. de la Place dans les Mémoires de l'académie 1784 et 1788, a fourni la forme des équations …"

As I previously showed the two-body problem had already been faced and solved by Newton in the hypothesis of two point-masses that attract each other according to the law of the inverse square of the distance. In our system centred on the Sun, the only one known at the time, there was another example where Newton's theory could be tested: it was the interaction between the Sun, Jupiter and Saturn, with the additional gravitational action exerted by the satellites of the two planets and by the other planets. However, it should be emphasized that the gravitational interaction between Sun-Jupiter-Saturn is very different from that between Sun-Earth-Moon.

Laplace had shown that all the inequalities of the motions of the Sun-Jupiter-Saturn system were the necessary results of a mutual action among celestial bodies according to Newton's theory of gravitation. The new Laplacian theoretical framework stimulated the calculation of new planetary and satellite tables. They were necessary because they provided information on which parameters had to be considered, and which had to be neglected in theoretical construction.

Although Laplace's research had prompted Delambre to study astronomy, he nevertheless claimed that the great French mathematician had prevented the Astronomical Observatory of Paris from establishing a systematic program of observations. In 1821 in an unpublished autobiography Delambre argued that it was not appropriate to put a

> … géometre à la tête d'un observatoire; il négligera toutes les observations, à la reserve de celle pour lequelle on aura besoin de des formules. (Bigourdan 1931, p. A.107)

The same criticism was directed at the Cassinis, who had never considered the drafting of a star catalogue which was considered the first and the more important thing to do in order to "…poser l'astronomie sur ses véritables fondemens … ." (Delambre 1810, p. 102)

Delambre's fame came essentially from his contribution to the definition of the Metric System, a problem closely related to the measurement of the meridian arc passing through Paris. Delambre's contributions were greatly appreciated not only by scientists but also by Napoleon to whom he had presented a copy of the text, in three volumes, which provided information on the operations that had been completed for the realization of the project. As Delambre himself wrote, the general would have told him: "Les conquêtes passent et ces operations restent." (Jacquinet 1949).

## 6.   Conclusions

## 6. Conclusions

From about 1740 onwards Cassini 1st semi-quantitative considerations on lunar libration were completely neglected, although his astronomical observations were taken into serious consideration. After Tisserand, who formalized Cassini 1st semi-quantitative considerations in three laws, dozens and dozens of articles are dedicated to them. They have become an object of contemporary research on which dozens of astronomers and mathematicians are engaged.

A detailed analysis of Cassini's contributions to astronomy is given by Delambre in *Astronomie moderne*, pp. 686-804. The overall judgment is negative but on lunar libration he credited Cassini 1st with stating that the Moon rotates around its axis. It is true that the rotation of the Moon is already found in Aristotle. And Hevelius also came very close to saying that the Moon has a rotational motion. But Delambre acknowledged that the rotation claim was clearly made by Cassini 1st and Cassini 2nd.

Cassini 1st had also assigned a numerical value - 2.5° - to the angle between the lunar equator and the plane of the orbit, a value that Kepler had left undetermined although he had highlighted its smallness. But the problem remained that Delambre had found no trace of Cassini's observations on lunar libration in the volumes of the Astronomical Observatory; a lack that Delambre considered very serious.

Delambre was scarcely a genius, at least as far as astronomy is concerned. As Wilson claims "he was, for his day, the applied scientist *par excellence*. … But [he] came at the right moment with appropriate skills and the necessary single-mindedness to lead the way in the working of a minor revolution." (Wilson 1980, pp. 269-270) He based his excellent redetermination of the orbital elements of the moving celestial bodies and in recalculation of astronomical tables on Laplace's theoretical discoveries and Maskelyne's observations.

In the last decade of his life, Delambre dedicated himself to the History of Astronomy which he retraced from antiquity to his own day. The text was aimed essentially at astronomers rather than historians.